\documentclass[12pt]{article}
\usepackage{graphicx,lscape,color,url,eurosym,hyperref}
\usepackage[a4paper, total={6.5in, 8in}]{geometry}
\usepackage[square]{natbib}
\usepackage{comment}

\title{Understanding Modern Banking Ledgers through Blockchain Technologies: Future of Transaction Processing and Smart Contracts on the Internet of Money}

\author{Gareth W. Peters$\ddag$ $\star$ $\ast$ \and Efstathios Panayi$\dag$ $\ast$\\
{\small{
$\ddag$ Department of Statistical Science, University College London}} \\
{\small{
$\star$ Associate Fellow, Oxford Mann Institute, Oxford University}}\\
{\small{
$\ast$ Associate Fellow, Systemic Risk Center, London School of Economics.}}\\
{\small{
$\dag$ UCL, Department of Computer Science, WC1E 6BT, London, UK}}\\
}

\begin{document}
\maketitle

\begin{abstract}
In this chapter we provide an overview of the concept of blockchain technology and its potential to disrupt the world of banking through facilitating global money remittance, smart contracts, automated banking ledgers and digital assets. In this regard, we first provide a brief overview of the core aspects of this technology, as well as the second-generation contract-based developments. From there we discuss key issues that must be considered in developing such ledger based technologies in a banking context.
\end{abstract}

%%%%%%%%%%%%%%%%%%%%%%%%%%%%%%%%%%%%%%%%%%%%%%%%%%%%%%%%
\section{Introduction}
%%%%%%%%%%%%%%%%%%%%%%%%%%%%%%%%%%%%%%%%%%%%%%%%%%%%%%%%

The 21st century has witnessed the emergence of a number of disruptive technologies. The advent of the internet has made the rise of \textit{social media} possible, and billions of people, as well as thousands of corporations are now connected and interacting on a daily basis. The world has also been able to leverage unused assets in the \textit{sharing economy}, which in many cases is powered by financial technology (fintech) firms, and this has also helped increase competitiveness in previously protected fields. \textit{Blockchain} holds promise for being the latest disruptive technology, and we argue in this chapter that the technology may find applications in areas as varied as transaction processing, government cash management, commercial bank ledger administration and clearing and settlement of financial assets. 

We have already seen how blockchain technologies have enabled the creation of crypto-currencies, and their rise has been documented widely\footnote{\cite{nakamoto2008bitcoin} introduced the first decentralised crypto-currency called Bitcoin, and related technology startups have already attracted more than \$1 billion in funding. The currency has been the subject of academic considerations in topics such as privacy (\cite{reid2013analysis}), security \cite{barber2012bitter}, regulation (\cite{peters2014opening}) and monetary policy (\cite{peters2015trends})}. The use of these currencies has prompted wide discussions on the merits (or lack thereof) of decentralisation, disintermediation, anonymity and censorship resistance in this setting, aspects which we discuss in this chapter also. One cannot, however, dispute their potential to disrupt areas such as the global remittance industry, by facilitating near-instantaneous global remittance with very low transaction fees.

In more recent times, blockchain applications have appeared that go far beyond their first application domains in virtual currencies, for instance they are now important in fields such as domain registration, crowdfunding, prediction markets and even gambling. Second generation blockchain technologies enable not only the execution of simple transactions, but the carrying out of computation on a network, where e.g. payments become conditional on the state of some internal or external variables (much the same way as financial derivatives have a payout that is a function of an underlying financial instrument). This is the basis for `smart contract' technologies, which we shall see can be important building blocks for these new application areas. As a consequence of these second generation technologies, a number of developments in this field have begun to appear which include third party data ledgers (\cite{devanbu2001authentic}), e-contracts/smart contracts and virtual contracts (\cite{buterin2014next,kosba2015hawk,swan2015blockchain}), e-assets or remote asset title transfers (\cite{halevi2011proofs}) and further applications, discussed in \cite{czepluch2015use}.

One can imagine that different applications require different blockchain structures or architectures, and our first contribution is in describing the differences (from both a theoretic and practical standpoint) between permissionless and permissioned blockchains in this context. As blockchains require nodes to act as verifiers for the network, permissionless blockchains allow for anyone to join as a verifier, while prior authorisation from a centralised authority or consortium is required in a permissioned blockchain. These blockchain types therefore require different approaches to achieving consensus, as well as incentivising verification activity on the network. In this context, one of our contributions in discussing these frameworks is to highlight the significance of data integrity protocols, which can be incorporated with blockchain technologies to achieve different levels of permissioning, data integrity and data security. 

While the potential for smart contracts to operate on these blockchains is very promising, it is not without its pitfalls. In particular, current blockchain structures, requiring the repetition of computation on all network nodes, will rapidly run into scalability issues, and this will require consideration before mass adoption becomes possible. Blockchain technology has the potential to revolutionize contract law and processing via self-enforcing digital contracts, whose execution does not require any human intervention. However, where automated smart contracts have a real-world counterpart, one has to understand both the legal and technical ramifications, particularly in the case of disagreements between the two, and we present current work in this area. 

In addition to these innovations in financial transactions and contract law, it is also recognised that such technologies can contribute significantly to other aspects of the financial industry, e.g. related to regulation and taxation. Importantly for this chapter, we envisage a banking and insurance environment in which blockhain technologies are utilised in banking ledger records and other banking and insurance records, such as loss databases and claims record databases. In this regard, we will discuss aspects that must be considered when developing such technologies for banking applications with regard to loss reporting, recording and provisioning in order to be consistent with modern regulations such as Basel III/IV, Solvency II and IFRS 9. 

We also discuss how blockchain technology also offers the potential for the development of new approaches to governance systems with the ability to decentralize many processes and thereby provide perhaps more democratic inclusive decision-making processes. Indeed there are already beginning to develop second generation blockchain architectures that are specifically designed for board rooms and automated structuring of governance frameworks for corporations. It is important to note what we refer to as decentralization in this context, see for instance the definition provided in \cite{benkler2006wealth}.
\begin{quotation}
Decentralization describes conditions under which the actions of many agents cohere and are effective despite the fact that they do not rely on reducing the number of people whose will counts to direct effective action.
\end{quotation}

Another area we explore in the context of blockchain technologies is that of clearing and settlement of financial assets. Several markets have experienced benefits in reducing counterparty and settlement risks in shortening the settlement cycle from 3 days to 2 days, and blockchain technologies have the potential to lead to near-instantaneous settlement. Because of the ability of blockchain to serve as an alternative for structures featuring centralised bodies for verification, we present the possible blockchain structures which would facilitate this, as well as initial industry attempts at pursuing this field.    

This chapter is topical, as it extends previous work regarding the real-world considerations that banks and other financial institutions would have, if they were to consider handling crypto-currency transactions (see, e.g. \cite{peters2014opening} for a discussion of operational risk considerations, and and \cite{peters2015trends} for a monetary policy perspective). In particular, it shifts the discussion to the underlying blockchain technology, which has a much broader scope for entering the banking sector and the regulatory space.

The remainder of the chapter is structured as follows: Section \ref{sec:blockintro} details the differences between the permissionless and permissioned blockchain types, it describes the advantages blockchains hold over databases and introduces smart contracts and their possible applications. Section \ref{sec:integrity} discusses existing notions of security, confidentiality, availability and integrity and how it applies to enterprise data, giving examples of blockchain structures which can preserve these features. Section \ref{sec:provisioning} proposes usecases for blockchain technology for government cash management through administering Treasury Single Accounts, as well as improving on the commercial bank ledger structures. Section \ref{sec:settlement} describes the current state of the clearing and settlement system and proposes blockchain approaches to reduce the inefficiencies. Section \ref{sec:conc} concludes.

%%%%%%%%%%%%%%%%%%%%%%%%%%%%%%%%%%%%%%%%%%%%%%%%%%%%%%%%
\section{Blockchain Technology Emerges}
\label{sec:blockintro}
%%%%%%%%%%%%%%%%%%%%%%%%%%%%%%%%%%%%%%%%%%%%%%%%%%%%%%%%

The emergence of blockchain technology is inextricably linked to the introduction of Bitcoin, the decentralised crypto-currency for the internet. \cite{nakamoto2008bitcoin} described how a network of users could engage in secure peer-to-peer financial transactions, eliminating the need for financial intermediaries and reducing the cost of overseas payments. In so doing, however, Nakamoto described a structure, termed the blockchain, along with a communication protocol, which essentially solved the Byzantine Generals' Problem\footnote{In the Byzantine Generals' problem, introduced by \cite{lamport1982byzantine}, a group of Byzantine Generals are camped around an enemy city in different locations. If they all attack simultaneously, then they have superior firepower to their enemy. The problem is that they need to agree a common battle plan, so that they attack at the same time, with the additional complication that there may be a traitor amongst their ranks.} and thus enabled the network to achieve consensus without requiring knowledge of users' identities, or trust relationships. 

So what is a blockchain and what are the different types of blockchain technology arising? In its most crude form, one may consider a blockchain to be a ledger or, more simply, a chronological database of transactions recorded by a network of computers. The term `blockchain' refers to these transactions being grouped in blocks, and the chain of these blocks forms the accepted history of transactions since the inception of the blockchain. On such a blockchain, anyone can attempt to provide an update to the blockchain ledger with a new record or amendment, which they sign with their own private cryptographic key. 

To ensure that only legitimate transactions are recorded into a blockchain, the network confirms that new transactions are valid, given the history of transactions recorded in previous blocks. A new block of data will be appended to the end of the blockchain only after the computers on the network reach consensus as to the validity of all the transactions that constitute it. Thus the transaction only becomes valid (`confirmed') once it is included in a block and published to the network. In this manner the blockchain protocols are able to ensure that transactions on a blockchain are valid and never recorded more than once, enabling people to coordinate individual transactions in a decentralized manner without the need to rely on a trusted authority to verify and clear all transactions.

%%%%%%%%%%%%%%%%%%%%%%%%%%%%%%%%%%%%%%%%%%%%%%%%%%%%%%%%
\subsection{Basics of Blockchain Technology}
%%%%%%%%%%%%%%%%%%%%%%%%%%%%%%%%%%%%%%%%%%%%%%%%%%%%%%%%

Just like many other technologies for the internet, blockchains rely on public key cryptography to protect users from having unauthorised persons take control of their accounts. The private and public key pairs enable people to encrypt information to transmit to each other, where the receiving party would then be able to determine whether the message actually originated from the right person, and whether it had been tampered with. This is critical when one needs to communicate to a network that a transaction between two parties has been agreed. In addition, the presence of an ability to identify the integrity of the data is also critical for applications we will consider as discussed further below. 

In this context, a useful concept is that of hash functions. The basic idea behind a hash functions use is to facilitate an efficient means for searching for data in a set of records. In its most basic form, a hash function is any function that can be used to map data of arbitrary size to data of fixed size where the output is a bit-string known as the hash value, hash code or hash sum. These hash values are stored in a tabular form known as the hash table and this is used as an efficient indexing mechanism when performing searches. Detailed discussions on the different types of hash function may be found in the overview of \cite{carter1977universal}

It turns out that when one combines hash functions and hash tables with cryptographic techniques, the resulting cryptographic hash function is directly applicable to establishing security and privacy protocols required for blockchain ledger technologies. In this context one can think of a cryptographic hash is like a signature for a text or a data file. It is secure since the cryptographic hash function allows one to easily verify that some input data maps to a given hash value, but if the input data is unknown, it is deliberately difficult to reconstruct it (or equivalent alternatives) by knowing the stored hash value. 

For instance, one of the most widely used blockchain technologies utilising a cryptographic hash function is the Bitcoin crypto currency protocol. In this instance it utilises a version of hash function with the following additional cryptographic features included, see discussion in Chapter 7 \cite{franco2014understanding}: One-wayness (i.e. premiage resistance) where knowledge of the hash value still makes it computationally highly improbable to find out the input data (a key element of proof-of-work aspects); weak and strong forms of collision resistance, the first of which means that given an input, it will be computationally improbable to find another input with the same hash value and stronger form states that it will be computationally improbable to find two input datas with the same hash value. Note, computationally improbable means here that no known algorithm can recover the input message from the hash within a time that is related polynomially related to the size of the input.
 
In the case of Bitcoin, this is achieved through the application of a SHA256 hashing function applied twice, see \cite{SHA256}. The SHA256 algorithm comes in several forms and is part of the SHA-2 class of hash functions, see discussions in \cite{SHA256,gilbert2004security,matusiewicz2005analysis}, but generally it generates an almost-unique, fixed size 256-bit (32-byte) hash security. Large classes of hash functions are based on a building block of a compression function, see discussions on this in \cite{merkle1980protocols,damgaard1990design,coron2005merkle}

Within the blockchain is also included information related to the digital time stamp, which records the temporal existence of a particular blockchain ledger item at a given instance in time. It could be utilised to symbolise that a contract between two agents is initiated or completed, that transactions of some form materialized or that payments/e-property were transferred ownership etc. Typically a digital time stamp also contains information relating to the hash created from the activity of securing the particular data/information entered into the ledger. This allows time stamping to occur with an element of privacy for the data being secured and entered on the blockchain ledger. In addition, just recording the hash is a more parsimonious representation of the information being secured or recorded. 

There exist parties such as Time-Stamping Authorities (TSA) that can provide a trusted third party arrangement to provide a secure and safe cold or secured live storage of information relating to the blockchain ledger recording. This digital notary signs with a private key for this data to be recorded and the time when this data was communicated to the authority. Then the signature address would be sent back to the original owner of the data. This simplified form is often performed in blockchain technologies using more advanced approaches such as a TSA collecting and securing in encrypted storage several agents data sets from within a fixed time period, then taking all data from this period and providing a time stamp, and hashing all this data together via a method such as a Merkle tree, see \cite{merkle1979secrecy,merkle1980protocols,devanbu2001authentic}. Then the resulting hash, for instance the root of the Merkle tree would be hashed together with the final hash of the previous time period and then published in the blockchain ledger. 

%%%%%%%%%%%%%%%%%%%%%%%%%%%%%%%%%%%%%%%%%%%%%%%%%%%%%%%%
\subsection{Permissionless and permissioned blockchains}
%%%%%%%%%%%%%%%%%%%%%%%%%%%%%%%%%%%%%%%%%%%%%%%%%%%%%%%%

There are various (often conflicting) categorisations of blockchain types, and for the purposes of this chapter we will focus on the different types of blockchain according to whether authorisation is required for network nodes which act as verifiers, and whether access to the blockchain data itself is public or private\footnote{\url{https://blog.ethereum.org/2015/08/07/on-public-and-private-blockchains/}}. For the first categorisation we have: 

\begin{itemize}
\item \textit{Permissionless} blockchains, where anyone can participate in the verification process, i.e. no prior authorisation is required and a user can contribute his/her computational power, usually in return for a monetary reward.  
\item \textit{Permissioned} blockchains, where verification nodes are preselected by a central authority or consortium.  
\end{itemize}

For the second categorisation we have:

\begin{itemize}
\item \textit{Public} blockchains, where anyone can read and submit transactions to the blockchain.
\item \textit{Private} blockchains, where this permission is restricted to users within an organisation or group of organisations.  
\end{itemize}

In reality, most permissionless blockchains feature public access, while the intention of most permissioned blockchains is to restrict data access to the company or consortium of companies that operate the blockchain. For this reason, we collapse the categorisation into two types, permissioned and permissionless blockchains, and we elaborate the distinction between them in the following section.  

%%%%%%%%%%%%%%%%%%%%%%%%%%%%%%%%%%%%%%%%%%%%%%%%%%%%%%%%
\subsubsection{Permissionless blockchains}
%%%%%%%%%%%%%%%%%%%%%%%%%%%%%%%%%%%%%%%%%%%%%%%%%%%%%%%%

In the prototypical example of a blockchain, the Bitcoin network, the blockchain used is `permissionless'. Permission refers to the authorisation for verification, and anybody can join the network to be a verifier without obtaining any prior permission to perform such network tasks. Because these verifiers are vital to the operation of the network, their participation is encouraged (and indeed incentivised) through the issuance of new currency that is paid to them once they have verified a block of transactions, the so called `Proof-of-Work' concept to be discussed below.

Consensus within the network is achieved through different voting mechanisms, the most common of which is Proof-of-Work, which depends on the amount of processing power donated to the network. The notion of Proof-of-Work allows the network to secure against malicious attempts to tamper with the blockchain structure due to the computational power that has already been applied to create the blockchain ledger entries. If an attacker wished to tamper with the blockchain, they would have to commit a computational effort equivalent or greater than all the power spent from the reference point they wished to alter to the present time. In addition, they would have to achieve this at a faster pace than the current legitimate network processing of new blockchain entries. Proof-of-work concepts can come in many forms, for instance they may rely on solutions to a computationally hard problem, a memory intensive problem or a problem that may require user interventions. To be practically useful for a blockchain technology, such problems must be computational challenging to solve, but efficient to verify a solution once obtained. Although these algorithms are vital in ensuring the security of the network, they are also very costly in terms of computation, and thus electricity usage also. 

A permissionless blockchain is advantageous in that it can \cite{swansonconsensus} both accommodate anonymous or `pseudonymous' actors and protect against a Sybil (i.e. identity-forging) attack \cite{douceur2002sybil}. On the other hand, the incentive mechanism has to be carefully developed in order to ensure that verifiers are incentivised to participate. In Bitcoin, for example, verifiers receive an amount for verifying each transaction, as well as for publishing a block of transactions. However, the latter is 2 orders of magnitude higher than the former. Since the incentive for publishing transactions of blocks decreases according to a predefined schedule, means that verifiers will at some point need to increase the amount they will require to process individual transactions, which makes it more costly to transact in Bitcoin. 

Besides Bitcoin, examples of permissionless blockchains include Ethereum\footnote{\url{https://www.ethereum.org}}, the platform that is intended to provide access to smart contracts on the blockchain, as well as offer blockchain as a service. 

%%%%%%%%%%%%%%%%%%%%%%%%%%%%%%%%%%%%%%%%%%%%%%%%%%%%%%%%
\subsubsection{Permissioned blockchains}
%%%%%%%%%%%%%%%%%%%%%%%%%%%%%%%%%%%%%%%%%%%%%%%%%%%%%%%%

This is not the only possible configuration of a blockchain, however, and the discussion is increasingly moving towards private, permissioned blockchains for specific usecases. Permissioned blockchains have a set of trusted parties to carry out verification, and additional verifiers can be added with the agreement of the current members or a central authority. Such a configuration is more similar to a traditional finance setting, which operates a Know Your Business (KYB) or Know Your Client (KYC) procedure to whitelist users that are allowed to undertake operations in a particular space. \cite{swansonconsensus} finds that permissionless and permissioned blockchains are fundamentally different in both their operation and the range of activities that they enable, some of which we review here. 

Permissioned blockchains are intended to be purpose-built, and can thus be created to maintain compatibility with existing applications (financial or otherwise). They can be fully private (i.e. where write permissions are kept within an organisation), or consortium blockchains (where the consensus process is controlled by a pre-selected set of nodes)\footnote{\url{https://blog.ethereum.org/2015/08/07/on-public-and-private-blockchains/}}. Because the actors on the network are named, the intention is that they are also legally accountable for their activity. In terms of the transactions these blockchains handle, it will be predominantly off-chain assets (such as digital representations of securities, fiat currencies and titles of ownership), rather than on-chain assets, such as virtual currency tokens \cite{swansonconsensus}. 

An advantage of a permissioned blockchain is scalability. In a permissionless blockchain, the data is stored on every computer in the network, and all nodes verify all transactions. It is obvious that once the number of transactions increases substantially, the users that are able to perform this type of processing and verification will decrease, leading to more centralisation. In a permissioned blockchain, only a smaller number of preselected participants will need to operate, and if these come from large institutions they will be able to scale their computing power in line with the increase in the number of transactions.  

However, because of the smaller number of participants, it is much easier for a group of users to collaborate and alter the rules, or revert transactions. In addition, it is easy for them to reject transactions and in this sense it is not `censorship resistant' as a permissionless blockchain would be. Examples of permissioned blockchains include Eris\footnote{\url{https://erisindustries.com/}}, Hyperledger\footnote{\url{http://hyperledger.com/}}, Ripple\footnote{\url{https://ripple.com/}} and others. 

%%%%%%%%%%%%%%%%%%%%%%%%%%%%%%%%%%%%%%%%%%%%%%%%%%%%%%%%
\subsubsection{Smart Contracts}
%%%%%%%%%%%%%%%%%%%%%%%%%%%%%%%%%%%%%%%%%%%%%%%%%%%%%%%%

In recent times, industry interest has increasingly moved to second generation blockhain applications, including digitising asset ownership, intellectual property and smart contracts. The latter usecase is particularly interesting, as one can encode the rules of a contract in computer code, which is replicated and executed across the blockchain's nodes. Such a contract can be self-enforcing, monitoring external inputs from trusted sources (e.g. the meteorological service, or a financial exchange) in order to settle according to the contract's stipulations. 

The concept of smart contracts has been considered as early as 20 years ago by \cite{szabo1997formalizing}, although we have only recently had concrete blockchain-based implementations. These blockchains extend the functionality of the network, enabling it to move from achieving consensus on data streams, to achieving consensus on computation (\cite{kosba2015hawk}). An example is Ethereum, which intends to provide ` built-in blockchain with a fully fledged Turing-complete programming language' (\cite{buterin2014next}).  

To understand how this extends the functionality of the Bitcoin blockchain, for example, let us consider a Bitcoin transaction as a very simple contract for the transfer of a certain amount of crypto-currency from one account to another. In a smart contract, this transfer could be made depending on some condition, for example: `Transfer x amount of y currency from Alice to Bob if the temperature in Devon is below 0 degrees Celsius on at least 20 of the next 30 days'. Smart contracts can feature loops and have internal state, so a much richer array of transactions becomes possible. In addition, they are permanent (i.e. they remain on the blockchain unless they are instructed to self-destroy), and are able to be reused as building blocks for a more complex service. 

Ethereum can be seen as a platform for deployment of internet services, for which such smart contracts are the building blocks. Because of the Turing-completeness of the in-built contract programming language, and the fact that computation is executed on every network node, it could have been possible for one to create an infinite loop, i.e. a contract that never terminates, which could bring down the network. To protect against this, programmable computation in Ethereum is funded by fees, termed `gas', and a transaction is considered invalid if a user's balance is insufficient to perform the associated computation (\cite{wood2014ethereum}).    

There are still potential issues to be resolved, however, before smart contracts can reach widespread adoption. One is scalability, as it is infeasible to expect that as the number of contracts and users grows, every single node has to process every transaction. The second is code correctness, as both the developers and users of the smart contracts have to be confident that the contract performs its intended use, and does not entail excessive fees due to unnecessary computations. Finally, there is the issue of the relationship between an electronic smart contract and its legal counterpart. How can one perform court enforced legally binding contracts on a distributed and decentralized system potentially over multiple legal jurisdictions. 

Thus far, smart contracts are not legally enforceable, although there have been efforts in the direction. Eris industries have recently proposed the idea of dual integration, or `ensuring a real world legal contract overlay fused onto a specific smart contract'\footnote{\url{https://erisindustries.com/components/erislegal/}}. Other initiatives include CommonAccord\footnote{\url{http://www.commonaccord.org/}}, which attempts to create templates of legal texts and thus create contracts in a modular fashion. The objective is to remove ambiguity as much as possible, having the smart contract accurately reflect the written legal contract, so that it can be actionable in the real world\footnote{\url{http://p2pfoundation.net/Legal_Framework_For_Crypto-Ledger_Transactions}}. 

%%%%%%%%%%%%%%%%%%%%%%%%%%%%%%%%%%%%%%%%%%%%%%%%%%%%%%%%
\subsection{Differences between blockchains and databases}
%%%%%%%%%%%%%%%%%%%%%%%%%%%%%%%%%%%%%%%%%%%%%%%%%%%%%%%%

In terms of applications of blockchain technology, one could argue that we are still in the exploration phase. It is prudent to be cautious about claims that this technology, particularly in its `permissioned blockchain' form could disrupt fields as diverse as banking, insurance, accounting etc. In particular, it would be useful to explore exactly what advantages blockchains have compared to well-understood transaction recording technologies, such as databases. 

To start with, we provide a short description about the types and capabilities of modern databases. Depending on the nature of the data one is storing, there are five genres of databases (\cite{redmond2012seven}):

\begin{itemize}
\item Relational databases, such as SQL and variants, which are based on set theory and implemented as two-dimensional tables;
\item Key-value stores, which store pairs of keys and values for fast retrieval;
\item Columnar databases, which store data in columns, and can have more efficient representations of sparse tables compared to relational databases;
\item Document databases; and
\item Graph databases, which model data as nodes and relationships.
\end{itemize}  

Databases can be centralised (residing at a single site) or distributed over many sites and connected by a computer network. We will focus on the latter, given the closer proximity to the blockchain concept. The objective of a distributed database is to partition larger information retrieval and processing problems into smaller ones, in order to be able to solve them more efficiently. In such databases, a user does not, as a general rule, need to be aware of the database network topology or the distribution of data across the different nodes. It should also be noted that in a distributed database, the connected nodes need not be homogeneous, in terms of the data that they store (\cite{elmasri2014fundamentals}). 

Because of the design of these databases and the replication of data across different nodes, such a database has several advantages (\cite{elmasri2014fundamentals} (p. 882)):
\begin{itemize}
\item Better reliability and availability, where localised faults do not make the system unavailable;
\item Improved performance / throughput;
\item Easier expansion.
\end{itemize}

In every distributed database, however, there is the issue of how modifications to the databases are propagated to the various nodes that should hold that data. The traditional approach is a `master-slave' relationship, where updates to a master database are then propagated to the various slaves. However, this means that the master database can become a bottleneck for performance. In multi-master replication\footnote{\url{http://www.multichain.com/blog/2015/07/bitcoin-vs-blockchain-debate/}} modifications can be made to any copy of the data, and then propagated to the others. There is a problem in this case also, when two copies of the data get modified by different write commands simultaneously.  

A blockchain could be seen as a new type of distributed database which can help prevent such conflicts. In the same way that the Bitcoin network will reject a transaction where the Bitcoin balance to be transferred has already been `spent', a blockchain can extend the operation of distributed databases by rejecting transactions which, e.g. delete a row that has already been deleted by a previous transaction (where a modification is a deletion, followed by the creation of a new row). 

A second difference between blockchains and distributed databases lies in the ability to create self-enforcing contracts that will modify the blockchain's data. Many permissioned blockchains have a built-in virtual machine, such that one can execute pieces of computer code on the network. If this virtual machine is Turing-complete, this means that the machine can potentially solve a very large set of problems, which is very useful for executing more complex transactions on the network, possibly conditional on the state of certain off-chain variables. 

The proliferation of databases as data stores has spawned considerations regarding data-related aspects, such as security, confidentiality and integrity. We argue that discussions around these issues will be important for blockchain technologies too, if they are to be successful in a business enterprise setting. In the following section we discuss these security aspects in depth and comment on blockchain attributes with regard to them.

%%%%%%%%%%%%%%%%%%%%%%%%%%%%%%%%%%%%%%%%%%%%%%%%%%%%%%%%
\section{Data security, confidentiality, availability and integrity on the blockchain}
\label{sec:integrity}
%%%%%%%%%%%%%%%%%%%%%%%%%%%%%%%%%%%%%%%%%%%%%%%%%%%%%%%%

Companies and organizations use the data they collect to personalize services, optimize the corporate decision making process, predict future trends and more. If one is to consider how to incorporate such records onto a blockchain there are fundamental issues to be considered. Within banking and financial services, for example, they must first of all adhere to different adopted best practices with regard to data confidentiality, availability and integrity. These concepts are related but distinct, and vital for data security within an organisation. We briefly discuss the first two, before exploring the latter in depth in the context of the blockchain.   

Confidentiality involves the protection of data from unauthorised disclosure, either by direct retrieval or by indirect logical inference. Confidentiality considerations can also involve the possibility that information may be disclosed by legitimate users acting as an information channel, passing secret information to unauthorised users. Within a blockchain, the choice of a permissioned or permissionless structure will define whether data will be made available to the public, or only within an organisation. Permissionless blockchains also enable carrying out transactions without the disclosure of private information. Because the operation of these blockchains rests on public key cryptography, securing users' private keys is critical, and indeed, this is one of the main source of operational risks in this area (\cite{peters2014opening}).  

Ensuring high availability means that data is accessible to authorised users. Availability is very closely related to integrity because service denial may cause or be caused by integrity violations. In blockchains, because data is replicated across many different nodes, availability should always be high. The catastrophic failure of a number of nodes should not cause any availability problems, although the network will experience a reduction in security proportional to the computational power of the missing nodes. 

%%%%%%%%%%%%%%%%%%%%%%%%%%%%%%%%%%%%%%%%%%%%%%%%%%%%%%%%
\subsection{Data integrity}
%%%%%%%%%%%%%%%%%%%%%%%%%%%%%%%%%%%%%%%%%%%%%%%%%%%%%%%%

Maintaining the integrity of data entails its protection from invalid modification, insertion or deletion, thereby preserving the accuracy, consistency and validity of data over its life cycle. Ensuring this integrity is important for the recoverability and searchability, as well as the traceability and connectivity of financial data records. This process usually requires a set of constraints or rules that define the correct states of a database or data set, and maintain correctness under operation. While the wider area of data security is often discussed in the context of cryptography considerations around blockchain technology, data integrity preservation in a blockchain structure has received comparatively little attention. 

When we discuss data integrity in this section we are referring to the accuracy and consistency or validity of data over its life cycle. In general, when discussing data integrity it may take one of two meanings, either referring to a state of the data or a to a process performed relating to the data. Data integrity in the context of a state specification defines a data set that is both valid and accurate, whereas data integrity as a process, describes protocols adopted to preserve validity and accuracy of a data set. 

In this section we argue that blockchain technologies can be structured to be consistent with best practices currently adopted with regard to data integrity, and indeed such considerations are only just starting to find their way into second generation blockchain technologies, with projects such as Enigma in MIT, see \cite{zyskind2015enigma}. The first aspect of data integrity we consider relates to the state of the data. In this context the state specification of the blockchain data records would be defined such that the data is valid and accurate or that the smart contracts operating on the blockchain are valid and accurate

The second aspect of data integrity relates to the transformation processes, which in this case would be operating on either data recorded or linked to the blockchain or to smart contracts operating on data on the blockchain. As with all applications in which data is a key ingredient, the data in its raw form is often not directly utilised for decision making and interpretation within the application. It must first undergo a variety of modifications and be put through different internal processes to transform the raw data forms to more usable formats that are practical for identifying relationships and facilitating informed decisions. Data integrity is then critical in ensuring that these transformations preserve the validity of the data set. Modern enterprises reliant on data, such as financial institutions, need to have confidence in the validity of their data, therefore they need to ensure both the provenance of the data and the preservation of integrity through transformation. Such transformations may also be included as part of smart contract structures in second generation blockchain applications.

Another way of conceptualising data integrity notions is to consider its function in preventing data modification by unauthorized parties, and maintaining internal and external consistency in the data set. If data is compromised then its utility to business practice and its informative nature may rapidly diminish. There are numerous places where such corruptions of data integrity may arise, for instance during replication or transfer or the execution of a smart contract that operates on a dataset to make contractual decisions. 

One can minimise such issues through the adoption of error checking methods. In a blockchain, these methods are inbuilt. The hash of each block of transactions is linked to the next, thus forming a chain. Transactions that are present in these blocks cannot be altered, unless one generates all blocks from that point onwards, which requires immense computational power. However, error checking for second generation blockchain structures, where smart contracts contain computer code, will be more involved.

Causes of corruption of data integrity include human error, which may or may not be intentional, code errors, viruses/malware, hacking, and other cyber threats, compromised hardware, such as a device or disk crash and the physical compromise to devices. From this list we see that some of these issues with preservation of data integrity rely on data security, whilst others are non-resolved when it comes to security solutions. We can easily see that data integrity and data security are related, and that data security refers to the protection of data against unauthorized access or corruption and is necessary to ensure data integrity. Hence, we see that data security is one of several measures which can be employed to maintain data integrity, as unauthorized access to sensitive data can lead to corruption or modification of records and data loss. 

In addition to security considerations, it is also clear that to achieve data integrity there is a strong case for data backup/ redundancy/ duplication. Often in practice, there are adopted best practices that remove the other data integrity concerns, such as input validation to preclude the entering of invalid data, error detection/data validation to identify errors in data transmission, and security measures such as data loss prevention, access control, data encryption.

We have already discussed the advantages that distributed databases bring in resolving a number of these issues. A blockchain additionally brings cryptographic security, a solution to the multi-master replication problem and facilitates more complex transactions through enabling smart contracts. 

\cite{ge2004secure} note the following aspects of integrity for data stored in databases which we believe would analogously apply to blockchain structures and architectures, as well as to perhaps different smart contract structures that are designed to operate on blockchain backbone networks:
\begin{itemize}
\item Integrity and consistency should involve semantic integrity constraints which are rules defining the
correct states of the system during operation. Such semantic constraints are present to ensure a level of automated protection against malicious or accidental modification of data, and ensure the logical consistency
of data. Rules can be defined on the static state of the database, or on transitions (as conditions to be verified before data is modified). In the context of blockchains, such rules would need to be applied at different levels, to the raw processing of data and in addition, to the functioning of smart contracts or secondary processes/transformations on blockchain related data records. 
\item Identification, Authentication, Audit. Before accessing a system, every user is identified and authenticated, both for the audit trail and for access permission. Auditing is the process of examining all security relevant events. Such features can be incorporated into blockchains either publicly through colouring mechanisms on the blockchain records (see overview in \cite{rosenfeld2012overview}) or through blockchain architectures such as permissioned blockchains.
\item Authorisation (access control). Authorisation applies a set of rules that defines who has what type of access to which information. Access control policies govern the disclosure and modification of information. In the context of requirement engineering, security aspects should not be afterthoughts of database design process and likewise should be included in many applications in the blockchain settings. Authorisations that are of relevance in blockchain settings can range from who can append data and what type to the blockchain, who can verify blockchain transactions, who can perform or execute smart contracts, view existing properties of smart contracts or initiate such smart contracts. 
\end{itemize}

In fact, the specification of protocols for preservation of data integrity in general governance structures has been a research topic since the early 70's and several different approaches have been proposed which include (not an exhaustive list):
\begin{enumerate}
\item The Bell LaPadula model (\cite{bell1973secure});
\item Discretionary Access Control protocols;
\item The Graham-Denning model (\cite{denning1976lattice}) and its extension the Harrison-Ruzzo-Ullman model (\cite{tripunitara2013foundational}). Note, the Graham-Denning Model is a computer security model that shows how subjects and objects should be securely created and deleted. It also addresses how to assign specific access rights. It is mainly used in access control mechanisms for distributed systems
\item Mandatory Access Control. Note, MAC often refers to an access control in which the operating system constrains the ability of a subject or initiator to access or generally perform some sort of operation on an object or target.
\item Multilevel security. Note, MLS usually involves the application of a computer system to process information with  different security levels, and prevents users from obtaining access to information for which they lack authorization. MLS is typically adopted in one of two settings, the first is to refer to a system that is adequate to protect itself from subversion and has robust mechanisms to separate information domains. The second context is to refer to an application of a computer that will require the computer to be strong enough to protect itself from subversion and possess adequate mechanisms to separate information domains, i.e a trusted system third party. 
\item The take-grant protection model. Note, this is a formal graphical model structure and protocol to establish or disprove the safety of a given computer system that follows specific rules. 
\item The Clark-Wilson model (\cite{clark1987comparison}).
\end{enumerate}

We briefly highlight a few examples of these data integrity frameworks mentioned and then comment on the ability to develop such frameworks within the context of a blockchain network. We specifically chose two frameworks, the Clark-Wilson model and the Biba model, as variants of them have wide spread uptake for the types of applications we will discuss.

%%%%%%%%%%%%%%%%%%%%%%%%%%%%%%%%%%%%%%%%%%%%%%%%%%%%%%%%
\subsection{Clark-Wilson Model for data integrity}
%%%%%%%%%%%%%%%%%%%%%%%%%%%%%%%%%%%%%%%%%%%%%%%%%%%%%%%%

We provide a brief review of the concepts behind the Clark-Wilson (CW) model which has been adopted in business and industry processes. \cite{clark1987comparison} argued the case for consideration of control over data integrity and not just considerations over control of disclosure. The Clark-Wilson model consists of subject/program/object triples and rules about data and application programs. The core of the CW model specification involves two key components: the notion of a transaction, which is characterized by a series of operations that transition a system from one consistent state to another consistent state; and the separation of duty (in banking settings often forming part of governance structures).

In the case of blockchain settings, one may think of the first component denoted the `transaction' as any function operating on the blockchain, such as addition of data to the blockchain, the verification of the blockchain transactions or a smart contract that reads or modifies the blockchain `state'. The second component involving separation of duty involves consideration of who may perform verification, who may view or alter data on the blockchain, or attached to the blockchain or who may view or execute or initiate such smart contracts.

In the CW model all data to be considered is partitioned into two sets termed Constrained Data Items (CDIs) and Unconstrained Data Items (UDIs). Then in addition, there are subjects which are entities that can apply transformation processes to data items to take CDIs from one valid state to another.  The term `transformational procedure' makes it clear that the program has integrity-relevance because it modifies or transforms data according to a rule or procedure. In addition, there are integrity validation procedures which confirm that all CDIs in a system satisfy a specified integrity scope. Data that transformational procedures modify are called CDIs because they are constrained in the sense that only transformational procedures may modify them and that integrity verification procedures exercise constraints on them to ensure that they have certain properties, of which consistency and conformance to the real world are two of the most significant. Then UDI's represent all other data, such as the keyed input to transformational procedures.

Given these structures, the CW model then specifies 6 basic rules that must be adhered to in order to maintain integrity of a system. We provide these below, along with a description of how these pertain to a blockchain structure.
\begin{itemize}
\item{The application of a transformation process to any CDI must maintain the integrity of the CDI and it may only be changed by a transformation process.

\textbf{Comment:} A transformation process is a transaction, and transactions on blockchains are unitary - it is impossible for one side of the transaction to happen without the other. Permissionless blockchains such as Bitcoin feature eventual consistency, in the sense that there may be some period of time for which the system is in an inconsistent state (e.g. due to a fork), but it is guaranteed to return to a consistent state eventually. However, this does however require consideration for structuring of smart contract and blockchain specific secondary application functions.

In permissioned/private blockchains, it is of possible to have near-instantaneous consistency.
}
\item{Only certain subjects can perform transformation processes on prespecified CDI's. This prespecification restriction must reflect governance enforced true separation of duties. The principle of separation of duty requires that the certifier of a transaction and the implementer be different entities.;

\textbf{Comment:} In any blockchain, subjects (users) are only enabled to transact with the tokens belonging to them, and no other user is able to access these without knowledge of their private key. Verifiers (miners) only ascertain whether transactions are valid. If however smart contracts have access to such information or data pertaining to users, this must be considered in their development on blockchain technology.
}
\item{All subjects in the system must be authenticated.;

\textbf{Comment:} This is the case in blockchain through public key cryptography.
}
\item{There must be a write only audit file that records all the transaction processes.;

\textbf{Comment:} Clearly this exists in the case of blockchain. An advantage of the blockchain is that it can provide guarantee of absence of modification. In the context of ownership, the blockchain proves that an asset has been transferred to somebody, and has not been transferred to somebody else subsequently. Because transactions can only be found on the blockchain, if a transaction is not found there, from the blockchain's perspective it does not exist\footnote{Factom whitepaper, available at \url{https://raw.githubusercontent.com/FactomProject/FactomDocs/master/Factom_Whitepaper.pdf}}.
}
\item{It must be possible to upgrade some UDI's to CDI's through the application of a transaction process.;
}
\item{Only a privileged subject in the system can alter the authorisations of subjects.

\textbf{Comment:} In the case of permissioned blockchains, there may be a consortium which may determine whether another node can enter the network. 
}
\end{itemize}

In most cases of blockchain architectures we have seen thus far the blockchains do not have hierarchies of authorisations for read and write access. The data model differs from that envisioned by Clark and Wilson, as there as many copies of the data as there are verifiers on the blockchain. Anybody can change their own data unilaterally, but for this to become the accepted history of the data, consensus has to be reached about the veracity of the new data. However, as people consider data integrity issues, there are emerging forms of blockchain architecture that will consider such features, such as the Enigma system developed in MIT, which will be discussed in this section.

Under the CW model, once subjects have been constrained so that they can gain access to objects only through specified transformational procedures, the transformational procedures can be embedded with whatever logic is needed to effect limitation of privilege and separation of duties. The transformational procedures can themselves control access of subjects to objects at a level of granularity finer than that available to the system. What is more, they can exercise finer controls (e.g., reasonableness and consistency checks on unconstrained data items) for such purposes as double-entry bookkeeping, thus making sure that whatever is subtracted from one account is added to another so that assets are conserved in transactions.

%%%%%%%%%%%%%%%%%%%%%%%%%%%%%%%%%%%%%%%%%%%%%%%%%%%%%%%%
\subsection{Biba models for data integrity}
%%%%%%%%%%%%%%%%%%%%%%%%%%%%%%%%%%%%%%%%%%%%%%%%%%%%%%%%

There are alternative approaches to considering data integrity, we also mention a second one that is an important contrast to the CW framework described above. The Biba models' (\cite{biba1977integrity}) conceptual framework deals primarily with integrity instead of confidentiality, with its key premise being that confidentiality and integrity are in concept the dual of each other, whereby confidentiality is a constraint on who can read a message, while conversely integrity is a constraint on who may have written or altered it. The design of this data integrity model is characterized by the phrase: `no read down, no write up'. Hence, in the Biba model, users can only create content at or below their own integrity level. Conversely, users can only view content at or above their own integrity level. This is in contrast to other integrity models, such as the Bell-LaPadula model, which is characterized by the phrase `no write down, no read up'. Extensions of the Biba model structure have also been explored in the Bell LaPadula frameworks, see for instance discussions on such lattice model structures in \cite{sandhu1993lattice} and \cite{ge2004secure}.

The Biba model is a formal state transition system of computer security policy that describes a set of access control rules designed to ensure data integrity. Data and subjects are grouped into ordered levels of integrity. The model is designed so that subjects may not corrupt objects in a level ranked higher than the subject, or be corrupted by objects from a lower level than the subject. Many applications therefore consider such models for data integrity as useful in for instance banking classification systems in order to prevent the untrusted modification of information and the tainting of information at higher classification levels.

The Biba model is summarised by the following three simple components:
\begin{itemize}
\item The subject should be able to read an object only if they have a higher than or equal security status than the object. This is known as the Simple Integrity Axiom and stated in another manner it says that a subject at a given level of integrity must not read an object at a lower integrity level (no read down).;
\item The subject should be able to write an object only if they have a lower than or equal security protocol than the object they write too. This is known as the Star Integrity Axiom and when stated in another manner it says that a subject at a given level of integrity must not write to any object at a higher level of integrity (no write up).; and
\item The final component is the Invocation Property which states that a process from a lower integrity level can not request higher integrity level access. In otherwords it can only interface with subjects at an equal or lower level of integrity status.
\end{itemize}

%%%%%%%%%%%%%%%%%%%%%%%%%%%%%%%%%%%%%%%%%%%%%%%%%%%%%%%%
\subsection{Integrity considerations in financial applications}
%%%%%%%%%%%%%%%%%%%%%%%%%%%%%%%%%%%%%%%%%%%%%%%%%%%%%%%%

There are blockchain technologies starting to emerge for financial applications which try to incorporate these notions of data integrity, security and confidentiality into their designs, we provide a few examples, though our coverage is by no means exhaustive in this rapidly growing field.

For instance in the case of crypto currency applications in blockchain technologies there is `cryptoassets.core' which includes `defensive programming principles to mitigate developer human error, data integrity and security issues'. In particular, in this context the irreversibility of the blockchain structure means one must be particularly cautious in financial processes to ensure that there are protocols to mitigate against human-errors by the developers and external attackers trying to exploit issues in the financial code. In this context, the following list of potential data integrity issues for cryptoasset services is provided\footnote{\url{http://cryptoassetscore.readthedocs.org/en/latest/integrity.html}}:
\begin{itemize}
\item Race conditions allowing over-balance withdrawals or account balance mix up (data integrity issues);
\item Double transaction broadcasts doing double withdrawals from hot wallet;
\item Partially lost data on unclean service shutdown;
\item Partially lost data when having Internet connectivity issues;
\item Database damage with bad migration;
\item Improper cold wallet handling increases the risk of losing customer assets;
\end{itemize}

In the developments considered in cryptoassets.core they are able to overcome issues such as race conditions through partitioned serialized transactions isolations. That is, the database transactions are performed in a complete isolation, one after each another, and thus there cannot be race conditions. If the database detects transactions touching the same data, only one of conflicting transactions may pass through and the others are aborted with application-level exception. In addition they build in features such as allowing each crypto currency/asset to obtain its own collection of database tables which contain ``static-typing like limits making it less likely for a developer to accidentally mix and match wrong...(assets)''. In addition they argue it is prudent to record in the blockchain transaction level time stamps for when it was broadcast to a network for verification or processing and a time stamp when this was completed, which a default of empty should it fail to complete. This allows for rapid verification and audit process for the blockchain to find non-processed transactions and then to recommit them for processing if they failed.

One of the disadvantages of using a blockchain as a data store in financial settings is that for the different verifying nodes to be able to agree on the blockchain's history, they have to store all of it locally in many blockchain architectures. While replication has benefits in increasing the resilience of the network against attacks, when the blockchain is predominantly used as a data store, it becomes more difficult to increase the storage capacity of all nodes as the amount of data increases. 

One solution to this, which maintains the cryptographic security of the data store has been proposed by MIT and their Enigma project (\cite{zyskind2015enigma}). This initiative has developed a decentralized blockchain-like framework utilizing three components: A blockchain, which records a complete history in append-only fashion, an off-chain distributed hash-table (DHT), which is accessible through the blockchain, and which stores references to the data, but not the data themselves, and a secure Multi Party Computation (MPC) component. Then any private data is encrypted on the client-side before storage and access-control protocols are programmed into the blockchain. 

Effectively, through this architecture the Enigma blockchain architecture provides a decentralized computation platform with guaranteed privacy that does not require a trusted third party's intervention. It achieves this through the use of secure multi-party computation and it ensures that data queries are computed in a distributed way, without a trusted third party. The actual data is partitioned over different nodes in the network, and they each compute functions together without leaking information to other nodes. Therefore, unlike blockchain approaches such as those created for Bitcoin, this blockchain has no network member ever having access to data in its entirety and instead each member has a seemingly meaningless chunk of the overall data. In addition, this lack of complete replication and reduction in redundancy can provide greater scalability features for the network and improve the speed. As noted in \cite{zyskind2015enigma} the `key new utility Enigma brings to the table is the ability to run computations on data, without having access to the raw data itself'. Clearly, a useful aspect for many financial processing applications of blockchain.

Basically, one can think of attaching Enigma to a particular type of blockchain architecture, and deploying it to enforce on the blockchain a feature of data integrity and privacy. It will connect to an existing blockchain and off-load private and intensive computations to an off-chain network. All transactions are facilitated by the blockchain, which enforces access-control based on digital signatures and programmable permissions. In this architecture the code to be executed (such as a smart contract etc) is performed both on the public blockchain and on Enigma for the private and computationally intensive components. It is argued that in this manner, Enigma can ensure both privacy and correctness. In addition, one can provide verified proofs of correctness on the blockchain for auditability purposes. 

Unlike blockchain approaches such as those underpinning bitcoin, in which execution in blockchains is decentralized but not distributed, meaning that every node redundantly executes common code and maintains a common state, in Enigma, the computational work is efficiently distributed across the network. 

As detailed in \cite{zyskind2015enigma} the off-chain network they develop in Enigma overcomes data integrity based issues that blockchain technology alone cannot handle as follows:
\begin{itemize}
\item The DHT, which is accessible through the blockchain, stores references to the data but not the data themselves. Private data is encrypted on the client-side before storage and access-control protocols are programmed into the blockchain.
\item It utilises privacy-enforcing computation in Enigma's network, in order to execute code without leaking the raw data to any of the nodes, while ensuring correct execution. This is key in replacing current centralised solutions and trusted overlay networks that process sensitive business logic in a way that negates the benefits of a blockchain. 
\end{itemize}

%%%%%%%%%%%%%%%%%%%%%%%%%%%%%%%%%%%%%%%%%%%%%%%%%%%%%%%%
\section{Considerations in Blockchain Technology Developments for Bank Ledgers and Financial Accounting}
\label{sec:provisioning}
%%%%%%%%%%%%%%%%%%%%%%%%%%%%%%%%%%%%%%%%%%%%%%%%%%%%%%%%

In this section we discuss items that may be of relevance in the banking and insurance sectors that could benefit from the developments of blockchain technologies. Industry publications, such as the recent Euro Banking Association report\footnote{Cryptotechnologies, a major IT innovation and catalyst for change, available at \url{https://www.abe-eba.eu/downloads/knowledge-and-research/EBA_20150511_EBA_Cryptotechnologies_a_major_IT_innovation_v1_0.pdf}} argue for the potential of blockchain technology to partly replace trusted third parties, commonly employed in many roles in finance as custodians, payment providers, poolers of risk and in insurance settings. Recall that the primary roles of such trusted third parties is to provide functionality such as: validation of trade transactions; prevention of duplicated transactions, the so-called `double-spending' issue; recording of transactions in the event of disputes over contract settlements or deliverables etc.; and acting as agents on behalf of associates or members. The blockchain can provide alternative solutions to fulfil these roles through the provision of a verifiable public record of all transactions which is distributed and can be decentralised in its administration.

In thinking about the possibilities of blockchain functionality within the banking sector, there are a range of different potential avenues to explore that move beyond the typical discussion on remittance services. Banking systems are large and complex, including a range of features such as back-end bookkeeping systems, which record customer account details, transaction processing systems, such as cash machine networks, all the way through to trading and sales, over the counter trades and interbank money transfer systems. To date, we are however unaware of any papers that go beyond this high level discussion and detail exactly how and what form blockchain technology may provide benefit in these aspects in banking settings.

In this work we will aim to provide a greater detail to these possible applications of blockchain technology. In particular we will discuss things related primarily to a few important unexplored areas: 
\begin{itemize}

\item Government cash management the central bank and treasury accounts in particular the Treasure Single Account (TSA) of \cite{pessoa2013government}; 

\item automation, decentralised and distributed banking ledgers; 

\item automated and distributed Over The Counter (OTC) contracts/products and clearing and settlement; 

\item automated client account reconciliations; and 

\item automated, distributed loss data reporting. 

\end{itemize}

One of the potential incentives for financial institutions, banks, insurers and banking regulators for the development of distributed blockchain technologies for these types of applications involves the reduction of overhead and costs associated with audit and regulation. In addition, more automation and efficiency in transaction processing, clearing and reconciliation can help to reduce counterparty credit risks.

Before entering into specific examples, we outline some core features that blockchain approaches share which can be both beneficial and detrimental. These require careful consideration when developing the applications to be discussed, as we have seen in previous discussions above on data integrity preservation.
\begin{itemize}
\item Immutability of itemized components in the blockchain. A blockchain is effectively a distributed transaction database or ledger that is immutable, in that data stored in the blockchain cannot be changed, i.e. deleted or modified. However, some versions of blockchain frameworks are starting to emerge that alter the perception of immutability, such as the Enigma project discussed above. This approaches the irreversibility of the standard blockchain framework by only allowing access to data for secure computations in reversible and controllable manners. In particular they also ensure that no one but the original data owner(s) ever see the raw data. 
\item Transparency of information presented on the blockchain. Many blockchains being created are publicly accessible by anyone with an internet connection and is replicated countless times on participating nodes in the network, though private versions or restricted blockchain networks are emerging also, as discussed previously. The question is to what extent the application requires private versus public components. In modern regulatory changes on banking and financial institutions there are numerous competing constraints which emphasize both the importance of financial disclosure, such as Pillar III of Basel III banking regulations, requiring financial institutions to demonstrate transparency in their reporting and their relationship with regulators, and on the other side there are also fiduciary duties that institutions maintain in upholding data privacy on behalf of their customers. Therefore, alternative approaches to private versus publick blockchain networks are also being explored where instead the data on a public ledger may have different levels of data integrity structure protocols which implement possibilities such as encryption of data stored in blockchains, see project Enigma for example (\cite{zyskind2015enigma}).
\end{itemize}

%%%%%%%%%%%%%%%%%%%%%%%%%%%%%%%%%%%%%%%%%%%%%%%%%%%%%%%%
\subsection{Possible Roles for Blockchain Technology Developments for Government Cash Management: Treasury Single Accounts}
%%%%%%%%%%%%%%%%%%%%%%%%%%%%%%%%%%%%%%%%%%%%%%%%%%%%%%%%

To understand how blockchain technology may benefit this application area to Government cash management we first need to describe briefly the concept of the treasury single account (TSA). The concept of TSA was first established widely in the consultative and technical notes of the authors based in the International Monetary Fund (IMF) by \cite{yaker2012treasury}\footnote{Note: the technical note is not representing the direct views of the IMF}.

The efficient functioning of a country or sovereign state hinges upon the efficiency of the government accounts. In this regard, government banking arrangements represent an important factor for efficient management and control of government’s cash resources. Therefore, the governments, central banks and countries treasury arrangements should be designed in such a fashion as to minimize the cost of government borrowing and maximize the opportunity cost of cash resources. The relationship between the treasury and the central bank is a core aspect of most financial policies in a country and is multifaceted in its components and working parts. A streamlined and coherent monetary policy structure in alignment with government financing policies and fiscal policy is a cornerstone of efficient macroeconomic strategy. One core principle to achieve this involves the basic step of ensuring that all cash received for projects operated or approved by government functions is available in a timely fashion for carrying out government's expenditure programs and making payments as required. It is particularly in this feature that we argue that blockchain architectures would benefit this application domain.

In this regard, a TSA is a single checking account for government funds from domestic revenue and some foreign funds (together called treasury funds) which are deposited and from which required money can be disbursed in timely fashion. It is a unified structure of government bank accounts that provides a consolidation of government cash resources in a common ledger. It was proposed to act as an essential tool for consolidating and managing governments’ cash resources, and in doing so it provided a means to reduce borrowing costs. This is particularly useful in countries with banking structures which are not well organised, fragmented and inefficient. Many countries have successfully established such TSA structure for their government accounts. Since the TSA structure is based on the principle of \cite{pessoa2013government} `unity of cash and the unity of treasury, a TSA is a bank account or a set of linked accounts through which the government transacts all its receipts and payments. The principle of unity follows from the fungibility of all cash irrespective of its end use'.

Several case studies now exist demonstrating the potential of such TSA account structures to improve the situation that many emerging market and low-income countries face relating to the fragmentation of their banking systems responsible for handling of government receipts and payments. Typically, in such countries that have not instigated such a TSA structure, there are significant challenges in the banking structures since the treasury may often lack governance and may not even possess a centralized control over the ruling government's cash resources. Consequently this can result in cash being idle or unavailable when required to fund core infrastructure projects or expenditure programs for development of the economy. It also extenuates the debt for such countries since the available cash for expenditures is laying idle for extended periods in numerous bank accounts that are fragmented and dispersed over different spending agencies, while in the meantime the government will be borrowing additional funds and increasing debt in order to mistakenly fund such projects and execute its budget plans. 

As detailed in \cite{pessoa2013government} if a country's government has such inefficiencies by lacking effective control over its cash resources there can be numerous consequences:
\begin{itemize}
\item Idle cash balances in bank accounts often fail to earn market-related remuneration.; 
\item The government, being unaware of these resources, incurs unnecessary borrowing costs on raising funds to cover a perceived cash shortage.;
\item Idle government cash balances in the commercial banking sector are not idle for the banks themselves, and can be used to extend credit. Draining this extra liquidity through open market operations also imposes costs on the central bank.
\end{itemize}
Clearly such inefficiencies can be improved with an automated trusted third party system such as that offered by some permission style blockchain technologies working as a ledger for such accounts under the auspice of a TSA structure.

Furthermore, one can argue for the TSA type structure since such a ``financial pool'' example is a concept that segregates two important aspects of this process, the separation of the finance function and the service delivery functions of sector offices and departments of regions, cantons and zones. We have seen that establishing such a feature in a blockchain would require application of data integrity functions. In concept such segregation of duties, is important as it allows public bodies to focus on their core responsibilities whilst financial execution such as control of cash, payments, accounting, reporting can be handled on their behalf by a pool of professionals or as we propose here, the automation by a privacy preserving, permissioned, distributed blockchain structure. Clearly in such a structure, public bodies still define and authorise their budgets and expenditures within these budgets.

%%%%%%%%%%%%%%%%%%%%%%%%%%%%%%%%%%%%%%%%%%%%%%%%%%%%%%%%
\subsubsection{A Complete Treasury Single Account Structure}
%%%%%%%%%%%%%%%%%%%%%%%%%%%%%%%%%%%%%%%%%%%%%%%%%%%%%%%%
Here we recall the definition provided in \cite{pessoa2013government} for a complete TSA structure which they argue has three core components:
\begin{enumerate}
\item Unification of the governments banking structure which will enable the treasury to have oversight of government cash flows in and out of these bank accounts and fungibility of all cash resources. Note particularly in a blockchain type extension can include a real-time electronic banking component and can also be private but remove the need for a single central oversight component in a government, instead replacing it with a distributed verification system.
\item In traditional TSA structures there would be no other government agency operating bank accounts outside the oversight of the treasury. In general the design of the TSA and access to the TSA will depend on the banking structures in place in a given country. In this regard, there are numerous different blockchain architectures that will be suitable for the blockchain version of the TSA. 
\item All government cash resources, both budgetary and extra-budgetary, should be included in the consolidated TSA account and the ``cash balance in the TSA main account is maintained at a level sufficient to meet the daily operational requirements of the government (sometimes together with an optional contingency, or buffer/reserve to meet unexpected fiscal volatility).'' (\cite{pessoa2013government}). Such features could be automated in blockchain versions of such an account through smart contract structures running on the blockchain TSA ledger. The minimum accounts that should be included in the TSA should cover all central government entities and their transactions. These include: 
\begin{itemize}
\item accounts managed by social security funds and other trust funds;
\item extra-budgetary funds (EBFs);
\item autonomous government entities; 
\item loans from multilateral institutions; and
\item donor aid resources; 
\end{itemize}
\end{enumerate}

The structure of the TSA ledger should contain accounts which include: 

\begin{itemize}
\item The treasury account, sometimes know as the TSA central account, which consolidates the governments cash position.;
\item TSA subsidiary accounts, which are netted off with the TSA main account, and which are typically created for accounting purposes in order aggregate a set of transactions whilst allowing the government to maintain the distinct accounting identity or ledger of its budget ministries and agencies.;
\item Transaction accounts for retail transaction banking operations which facilitate access indirectly to the TSA for different government projects. Typically such accounts are either imprest accounts (accounts with a cash balance which is capped) or zero balance meaning that their cash balances are aggregated back to the TSA main account on a regular schedule, such as daily. Such zero balance accounts are advantageous as they do not interact with interbank settlement processes for each transaction, further improving efficiency and access to funds.;
\item  Transit accounts for flows of cash into and out of the TSA main account.; and
\item Correspondent accounts. The development of daily clearing/netting can be automated with smart contract structures on a TSA banking ledger blockchain.
\end{itemize}

%%%%%%%%%%%%%%%%%%%%%%%%%%%%%%%%%%%%%%%%%%%%%%%%%%%%%%%%
\subsubsection{Possible Blockchain Architectures for a Complete TSA}
%%%%%%%%%%%%%%%%%%%%%%%%%%%%%%%%%%%%%%%%%%%%%%%%%%%%%%%%
Note in a standard TSA banking structure there are numerous possibilities relating to where the TSA account should be maintained. That is in which institution, in most cases it is argued that the Central bank of the country may be the appropriate venue for such an account. However, all cases require a trusted third party to administer and provide governance and oversight of this critical consolidated account. One potential advantage of a TSA account placed under a blockchain structure with access through private key permissions and smart contracts is that when it is placed under a private network, there would not need to be a single point of administration, it would be a trusted closed network which has been effectively decentralized. We argue that this could have the potential to remove issues that may arise with governance and oversight when entrusting such key accounts to one single institution, especially in a fragmented banking and institutional structure. 

In addition, as noted in \cite{pessoa2013government}, one can establish the TSA banking structure with ledger sub-accounts in a single banking institution (not necessarily a central bank), and can accommodate external zero balance accounts (ZBAs) in a number of commercial banks. This can still be achieved in a blockchain framework in several different architectures. One of which may involve separate ZBA's maintained as internal blockchain ledgers in each commercial bank in the network, followed by a linking of these blockchains to the main TSA blockchain ledger account, with links being executed for transactions based on smart contract technology.

There are two main categories for architecture for a TSA banking account which align well with different architectural decisions for the blockchain version of the TSA, these involve either a centralized or a distributed architecture. In the centralized structure, all revenue and expenditures of the government are on a single ledger which is administered by a single trusted third party such as the central bank, whereas, in the distributed TSA structure there will be a hierarchy of organizational structures each with their own separate transaction accounts in the banking system, but still a single TSA account that contains all balances by close of business each day. As detailed in \cite{pessoa2013government}, the Sweedish TSA structure involves zero balance accounts in the central bank which are authorised by the minister for finance and available for individual spending agencies. In this case, money is transferred from the TSA to such accounts for payment of authorised project expenditures as required. All these accounts are cleared/reconciled with the TSA account daily.

In addition to understanding the network structure and components of the TSA account, there is also the aspect of revenue collection, remittance, payments and processing under the TSA account to be discussed and to consider how blockchain technology can facilitate the functioning, automation and decentralization of such important components. In traditional TSA structures there are two possibilities as detailed in \cite[Section II Part A]{pessoa2013government} where they describe either centralized transaction processing or decentralized approaches. Under the blockchain version of such a TSA structure, all transactions processing and remittance could be performed in a decentralized manner which would not require oversight of a particular trusted third party network member, as all transaction processing would be automated on the blockchain.

Futhermore, such automation of these processes into the blockchain structure would be more efficient and less costly than existing practices. Currently the best practice adopted for TSA account processing is to utilise the commercial banking network. That is, it is common to contract the commercial banks for revenue collection purposes, these commercial banks transfer revenues collected to the TSA main account daily (so as to avoid float or need of imprest cash amounts). As noted in \cite{pessoa2013government} such a system, though widely utilised, involves a remuneration system which ``...is not transparent and does not clearly indicate the cost of revenue collection services provided by banks. The banks use the free float to invest in interest-bearing securities. This process clearly distorts the TSA structure and concept''.

In this manner one could argue that the blockchain version of the TSA account structure can further reduce costs associated with remittance services and revenue collection services currently charged to the government on a fixed contractual basis. It is currently the case that such remittance services, when provided in a blockchain structure such as in the bitcoin network, can be orders of magnitude cheaper than those offered by traditional remittance providers in financial services industries.

%%%%%%%%%%%%%%%%%%%%%%%%%%%%%%%%%%%%%%%%%%%%%%%%%%%%%%%%
\subsection{Possible Roles for Blockchain Technology Developments for Commercial Bank Ledgers}
%%%%%%%%%%%%%%%%%%%%%%%%%%%%%%%%%%%%%%%%%%%%%%%%%%%%%%%%

Here we discuss considerations around blockchain technologies when they are utilised for retail and investment banking account ledgers. If one is to develop blockchain technologies for ledgers in banks, it is first important to understand some primary components of typical ledgers that will need to be hashed into the blockchain structure and considered in designing the blockchain architectures and the data integrity and governance functions. Hence, we provide a simple overview of the basic structure of a banking ledger in a modern large scale retail bank. One can think of the ledger accounts of a business as the main source of information used to prepare all their financial statements and reporting. In standard practice, if a business were to update their ledgers each time a transaction occurred, the ledger accounts would quickly become cluttered and errors might be made. This would also be a very time consuming process, but this may change with the new potential of blockchain technologies. Before considerations of this nature, we overview the standard approach and then discuss some aspects.

A typical banking system has a range of different structures, including the account master file containing all customers' current account balances together with previous transactions for a specified period, such as 90 days.; a number of ledgers in place to track cash and assets through the processing system; a variety of journals to hold transactions that have been received from a range of different sources not yet entered in ledgers; and an audit trail which records who did what, when they did it and where they did it regarding transaction processing. There will also be a suite of processing software that will operate on this data for overnight batch processing to apply transactions from journals to various ledgers and account master file updating. In addition, there are audit processes in place to scrutinize the functioning of all these systems. In automated systems there are different approaches, one developed in 1987 and widely adopted is the Clarke-Wilson Security Policy Model, see details in \cite{clark1987comparison} and the brief description made above in the section on data integrity. Such an approach sets out how the data items in the system should be kept valid from one state of the system to the next and specifies the capabilities of various principals in the system.
 
In addition to the component structures in the banking ledgers, there are also processes to understand in how these ledgers are utilised. Therefore, to understand the ledger process better we must first recall some basics of the financial reporting cycle. In the case of the financial cycle, the accounting system records two economic events: the acquisition of capital from creditors and owners and the use of that capital to acquire property to generate revenue. This cycle performs also a third function - the financial reporting function. 

We first note that the word `ledger' refers to a book or set of records. In general, when working towards construction of the banking ledger, initially all transactions are recorded in what are known as the books/ledgers of prime entry. These are records of the transaction, the relevant customer/supplier and the amount of the transaction, which is essentially a daily list of transactions. There are a range of different prime entry books which typically consist of different types of transactions: 
\begin{itemize}
\item Sales day book which records credit sales; 
\item Purchase day book which records credit purchases; 
\item Sales returns day book which records returns of goods sold on credit; 
\item Purchase returns day book which records goods returned bought on credit; 
\item Cash book which records all bank transactions;
\item Petty cash book which records all small cash transactions; and the
\item Journal which records all transactions not already recorded.
\end{itemize}

In general, the prime entry books/ledgers serve to `capture' transactions as soon as possible so that they are not subsequently lost or forgotten about. The cash book and the petty cash book are part of the double entry system and record cash coming in and going out. Note, double entry accounting systems are just those that record transactions in two different separate books, as a credit in one and a debit in the other. However, the day books and journal are not part of the double entry system, and entries are made from there to the ledgers. Then all these prime entry books are summarised and the total of the daily transactions is recorded in the accounting ledgers of the company, which is done in the standard `double entry' format. 

After the prime entry books/ledgers there are other key elements of the double entry system in financial accounting. This system consists of the three basic components: the general ledger, the cash book and the petty cash book. In addition, there are the receivables and payables ledgers to be considered which provide details of the total receivables and payables that are recorded in the nominal ledger. 

The general or nominal ledger is responsible for the recording of all accounts such as wages, sales, purchases, electricity, travel, advertising, rent, insurance, repairs, receivables, payables and non-current assets. In addition, one typically in principle considers the cash and bank accounts as forming part of this ledger, however they are typically physically recorded in a separate book since the number of such transactions is usually very large for some businesses.

The total amounts owed to suppliers and to whom it is owed is recorded in the payables ledger. Typically, there would be a separate account maintained in the ledger for each individual supplier. Generally, the aggregate total of the outstanding amounts owed in the payable ledger should reconcile with the payables balance in the general ledger. Conversely, to record the total amount owed to the business by customers there is the analogous ledger often called the receivables book. It provides details of exactly what is owed and from whom, as with the payables ledger, it is common practice to have a separate account for each customer. Again, there should be a reconciliation between the total amounts owing in the receivables ledger and the receivables balance in the general ledger.

In the context of a banking financial institution, there is a primary concern regarding the debt and equity capital records. These are the sources of such an organization’s capital which includes its creditors and owners. Such institutions typically distinguish three forms of capital transactions: 
\begin{enumerate}
\item Bank Loans which are reported in the Notes Payable account and are to be utilised for short, medium or long term financing and typically asset backed. The Notes Payable Ledgers record outstanding notes and include aspects such as the identifier for the note, its holder, its maturity and current interest rate, and the original and current balance.
\item Bond Issuances are summarized in the Bonds Payable balance sheet item and specific details of each issuance are detailed in a journal entry record for each with individual separate long-term liability accounts. Such issuances allow financial institutions to obtain medium and long term capital, but they create a contractual obligation to pay a fixed amount of interest at specified intervals in the future. 
\item Stock Issues are recorded as par or stated value for issued shares in Common Stock and Preferred Stock account ledger. Typically, a corporation will use different Capital Stock accounts for each class of stock.
\end{enumerate}

Apart from the debt and equity capital recording ledgers, there are also other key ledgers to consider such as the Property and Systems records. These keep track of all depreciating property, plant and equipment. Typically it would involve the recording of three primary types of transactions which include:
\begin{enumerate}
\item Acquisition of property where an institution would use buildings and equipment to generate revenue;
\item Depreciation of property, plant and equipment; and
\item Disposition of property.
\end{enumerate}

To complete the third key component of financial reporting systems we also mention the basic idea of the Journal Entry and Financial Reporting Systems where insitutions are able to record transaction in the general ledger using three types of accounting entries that summarize `High-Volume' transactions such as sales and purchases, `Low-Volume' transactions such as changes in debt and equity capital in order to remove depreciating property etc., and closing entries.

In the standard ledger systems described above different governance structures may be put in place to resolve potential accounting malpractice and fraud from arising. This is typically achieved by separating responsibilities for the double entry process, otherwise known as dual control systems or decentralized responsibility. These are typically enactments of different data integrity processes as described previously.

We note that with the development of blockchain ledger technologies and smart contracts, many of these processes and ledgers/accounting books just described which make up the financial accounting system can be automated through a blockchain structure. Such blockchains could take many forms, they could be either distributed within an organisation or even available as a public ledger (perhaps with some form of encryption for private data) that is shared between institutions, regulators and government agencies undertaking oversight, taxation etc.

In developing blockchain ledgers, one will need to consider how often to consider constructing hash entries, i.e. how often to aggregate data together before adding a hash entry to the blockchain ledger, as described in the previous overview section. In the context of banking ledgers, this will probably depend on the types of data being placed on the ledger, for instance, one may consider such data as divided into the following components. Transaction data is one source of data that must be considered, it is the records of information about each transaction and it is expected to change regularly as transactions progress and are completed. Other less frequent data includes aspects of standing or reference or meta data which is typically for all practical purposes permanent and includes names and addresses, descriptions and prices of products. We would suggest that the blockchain hashing for such financial records and ledger creation should follow a combination of both batch processing and in some applications real-time processing.

There are a few developments of second generation blockchain ledgers being developed for such banking ledger processing. For instance Balanc3\footnote{\url{https://consensys.net/ventures/spokes/}} is a new blockchain technology being developed based on smart contracts and a blockchain architecture with the purpose of performing acounting ledger processing, in this case as a triple entry system, rather than the double entry systems described above. It is argued that the non-repudiability and comprehensive audibility of the blockchain can be utilised to guarantee the integrity of accounting records. The blockchain architecture in this case utilises a range of different products for data integrity and security, they include EtherSign combined with an IPFS decentralized data storage platform and smart contracts enacted through Ethereum blockchain. In this way this account ledger is able to construct, store, manage, and digitally sign documents. The admissible documents in this system include features such as self-enforcing smart contracts like employment contracts or invoices, or traditional text agreements. Invoicing through smart contracts automatically processes and records payments. 

Before exploring other applications of blockchain, it is worth to observe that the immutability of the blockchain record when automated for transactions must be carefully considered in some cases. For instance, the issue of loss provisioning in the ledger of a banking institution must be carefully considered. We briefly comment on this below.

%%%%%%%%%%%%%%%%%%%%%%%%%%%%%%%%%%%%%%%%%%%%%%%%%%%%%%%%
\subsubsection{Considerations of Blockchain Ledgers Regarding automated Provisioning Processes of Losses Under IFRS9}
%%%%%%%%%%%%%%%%%%%%%%%%%%%%%%%%%%%%%%%%%%%%%%%%%%%%%%%%
After the 2008 financial crisis that affected the world wide banking sector there were significant changes to banking regulations, insurance regulations and accounting standards. In particular the regulation we mention in this section that must be carefully considered in blockchain ledger applications in banks is that of the provisioning accounting standard now known as IFRS 9 Financial Instruments document, published in July 2014 by the International Accounting Standards Board (IASB)\footnote{\url{http://www.ifrs.org/current-projects/iasb-projects/financial-instruments-a-replacement-of-ias-39-financial-instruments-recognitio/Pages/Financial-Instruments-Replacement-of-IAS-39.aspx}}. 

This standard has three core components:
\begin{itemize}

\item Classification and measurement. Any blockchain based ledger for banking settings must ensure that its recording and processing of losses in the ledger are compliant with the IFRS 9 standard classifications. These classifications determine how financial assets and financial liabilities are accounted for in financial statements and importantly how they are measured on an ongoing basis. This is particularly relevant in some blockchain architectures in which immutability would not allow reversal of misclassifications. Under the IFRS 9 framework, there is a logical classification structure for all financial assets which is driven by cash flow characteristics and the business model in which an asset is held. Under such a framework, it standardizes the management and reporting of such items in banking ledgers in a simplified fashion compared to those of the previous rule-based requirements that are complex and difficult to apply, and more importantly, complicated to automate into smart contract transformations on a blockchain banking ledger. 

\item Impairment components relate to the management of delayed recognition of credit losses on loans and other financial instruments. Under this aspect of IFRS 9, there is a new expected loss impairment model that will need to be enacted via smart contracts on the blockchain banking ledger processing system. This process would provide an automated and timely recognition of expected credit losses. The new Standard requires entities to account for expected credit losses from when financial instruments are first recognised, so this must be incorporated into the banking ledger processing and could be done in an automated manner with a model enacted via a smart contract structuring. In addition, the IFRS 9 standard makes impairment easier to deal with in an automated smart contract structure, as it is much more standardised under IFRS 9, making it also easier to develop such features in a unified fashion for all financial instruments. This standardization across all financial assets removes a source of complexity associated with previous accounting requirements.

\item The third aspect of IFRS 9 relates to hedging accounting changes.

\end{itemize}

IFRS 9 applies one classification approach for all types of financial assets, including those that contain embedded derivative features, making the design of smart contracts based on such classifications much more streamlined and simpler to automate. Under IFRS 9, one has four possible classification categories for financial assets, which are now classified in their entirety rather than being subject to complex bifurcation requirements, again making things much more streamlined for smart contract based automations. The classification process can be potentially automated via a smart contract structure, see page 7 process model for such classification automation \footnote{\url{http://www.ifrs.org/current-projects/iasb-projects/financial-instruments-a-replacement-of-ias-39-financial-instruments-recognitio/documents/ifrs-9-project-summary-july-2014.pdf}}.

In addition, the smart contract implementation would need to be developed under IFRS 9 compliance to apply two criteria to determine how financial assets should be classified and measured: the entity's business model for managing the financial assets; and the contractual cash flow characteristics of the financial asset. It should be noted that unless an asset being classified meets both test requirements, then it will be recorded on the blockchain banking ledger in terms of fair value reporting in the profit and loss. If the asset passes the contractual cash flows test, the business model assessment determines how the instrument is classified. For instance, if the instrument is being held to collect contractual cash flows only then it is classified as amortized cost. However, if the instrument is to both collect contractual cash flows and potentially sell the asset, it is reported at fair value through other comprehensive income (FVOCI). Further background on the requirements of the IFRS 9 provisioning standards that blockchain ledgers and processing must adhere to can be found at IASB documents, \url{http://www.ifrs.org}.

In addition to the role of blockchain in banking ledgers, blockchain and smart contracts may also be utilised for other roles, such as the clearing and settlement processes. We briefly outline the role of settlement processes in the following section.

%%%%%%%%%%%%%%%%%%%%%%%%%%%%%%%%%%%%%%%%%%%%%%%%%%%%%%%%%%%%%%%%%%%%%%%%%%%%%%%%%%%%%%%%%%%%%%%%%%%%%%%%%%%%%%%%
\section{Blockchain Technology and the Trade Settlement Process}
\label{sec:settlement}
%%%%%%%%%%%%%%%%%%%%%%%%%%%%%%%%%%%%%%%%%%%%%%%%%%%%%%%%%%%%%%%%%%%%%%%%%%%%%%%%%%%%%%%%%%%%%%%%%%%%%%%%%%%%%%%%

A banking area which has been hampered by the inefficiencies of traditional processes is that of settlement of financial assets. Major markets such as the US, Canada and Japan still have a 3-day settlement cycle (T+3) in place, while the EU, Hong Kong and South Korea have moved to T+2. This delay in settlement drives a number of risks, which we will discuss in the following section.

In order to understand how blockchain technology could potentially enter into this field, it is useful to overview the lifecycle of a trade. Firstly, a buyer comes to an agreement with a seller for the purchase of a security. What follows then is referred to as clearing, when the two counterparties update their accounts and arrange for the transfer of the security and the associated monies. This process entails\footnote{Source: Risk management issues in central counterparty series, presentation by Priyanka Malhotra at the Systemic Risk Centre at LSE.}:
\begin{itemize}
\item Trade valuation;
\item Credit monitoring;
\item Position management;
\item Member reporting;
\item Risk management;
\item Collateral management;
\item Netting of trades to single positions;
\item Tax handling;
\item Failure handling.
\end{itemize}

The actual exchange of the money and securities is termed settlement, and completes the cycle - this is typically 2 or 3 days after trade execution and can involve the services of institutions, such as custodians, transfer agents, and others (\cite{bliss2006derivatives}). In a typical trading-clearing-settlement cycle, the following actors are involved:
\begin{itemize}
\item On the trading side
\begin{itemize}
\item The investors (buyer and seller) who wish to trade
\item Trading members (one for the buyer and one for the seller) through which the investors can place their orders on the exchange
\item The financial exchange or multilateral trading facility, where the trading members place the trades
\end{itemize}
\item On the clearing side
\begin{itemize}
\item The clearing members, who have access to the clearing house in order to settle trades. These firms are also trading members, and thus settle their own trades, but non-clearing trading members have to settle through them. 
\item The clearing house / CCP, which stands between two clearing members.  
\end{itemize}
\item On the settlement side
\begin{itemize}
\item The two custodians, who are responsible for safeguarding the investors' assets. This role may also be played by a Central Securities Depository (CSD)
\item The settlement system.
\end{itemize}
\end{itemize}

Clearing and settlement can be bilateral, i.e. settled by the parties to each contract. This of course entails that risk management practices, such as collateralisation, are also dealt with bilaterally (\cite{bliss2006derivatives}). A large number of OTC (over-the-counter) derivatives used to be settled in this way, until recent efforts by the G20 resulted in regulation to impose central clearing \footnote{For an example of EU regulation in this area, see \url{http://www.fca.org.uk/firms/being-regulated/meeting-your-obligations/firm-guides/emir}.}.   

In central clearing, a third actor acts as a counterparty for the two parties in the contract and is termed the central counterparty (CCP). This simplifies the risk management process, as firms now have a single counterparty to their transactions. Through a process termed novation, the CCP enters into bilateral contracts with the two counterparties, and these contract essentially replace what would have been a single contract in the bilateral clearing case. This also leads to some contract standardisation. In addition, there is a general reduction in the capital required, due to multilateral netting of cash and fungible securities. \cite{duffie2011does} discusses the effects of introducing a CCP for a particular class of derivatives. Examples of securities that are centrally cleared include equities, commodities, bonds, swaps, repos etc.

There are two main risks that are exacerbated by a longer settlement cycle:
\begin{itemize}
\item Counterparty risk between trade execution and settlement, and associated margin requirements. Because of these risks, clearing members are required to maintain capital with the CCP. One can see that faster execution times would minimise this risk, and thus minimise these capital requirements. As an example of this reduction, note that a `move from T+3 to T+2 implies a 15\% and 24\% reduction in the average Clearing Fund amount, during the typical and high volatility periods, respectively'\footnote{BCG, Shortening the Settlement Cycle October 2012, available at \url{http://www.dtcc.com/~/media/Files/Downloads/WhitePapers/CBA_BCG_Shortening_the_Settlement_Cycle_October2012.pd.f}}.
\item Settlement risk, or `the risk that one leg of the transaction may be completed but not the other'(\cite{bliss2006derivatives}). Certain payment methods have been proposed to combat this issue, but it is obvious that a shorter settlement cycle would mitigate this further. 
\end{itemize}

The two main types of settlement instructions are Delivery vs Payment (DvP) and Free of Payment (FoP), also termed Delivery vs Free \footnote{\url{https://www.fanniemae.com/content/fact_sheet/dvp-dvf-comparison.pdf}}. The former ensures that delivery of the assets will only occur if the associated payment occurs. The latter method is simply a free delivery of assets, i.e. it is not associated with a payment, which occurs in a separate transaction. This obviously introduces a risk of non-payment by the buyer. 

We next describe the possible blockchain structures that could be used to mitigate risks and increase efficiency. 

%%%%%%%%%%%%%%%%%%%%%%%%%%%%%%%%%%%%%%%%%%%%%%%%%%%%%%%%
%%%%%%%%%%%%%%%%%%%%%%%%%%%%%%%%%%%%%%%%%%%%%%%%%%%%%%%%
\subsection{Blockchain in the settlement cycle}
%%%%%%%%%%%%%%%%%%%%%%%%%%%%%%%%%%%%%%%%%%%%%%%%%%%%%%%%
%%%%%%%%%%%%%%%%%%%%%%%%%%%%%%%%%%%%%%%%%%%%%%%%%%%%%%%%

In most applications of the blockchain, the advantages are that it brings are decentralisation and disintermediation. In the case of the trading-clearing-settlement cycle, we can envision the possibility of a consortium blockchain used at every level:

\begin{itemize}
\item At the financial exchange / multilateral trading facility level. For example, a consortium of brokers can set up a distributed exchange, where each of them operate a node to validate transactions. The investors still trade through a broker (due to naked access regulations), but the exchange fees can be drastically reduced. 
\item At the clearing level. A consortium of clearing members can set up a distributed clearing house, thus eliminating the need for a CCP. Clearing then becomes closer to bilateral clearing, with the difference that the contract stipulations are administered through a smart contract, and thus there is less scope for risk management issues.  
\item At the settlement / custodian level. 
\end{itemize}

A concrete example of how the entire lifecycle of a trade would look like is as follows: A buyer submits an order to buy a particular amount of an asset, for which there is an equivalent selling interest, through his broker. The buyer's and seller's brokers then create a transaction for the transfer of that amount of the asset, which is then transmitted to the distributed exchange network and verified. Once a block of transactions is verified, it is transmitted to the decentralised clearing house, where a new transaction is created, involving the brokers' clearing members. Once this transaction is verified in the clearing house blockchain, it is then transmitted to the settlement system, where a new transaction is created involving the custodians or CSDs, and the transfer of assets occurs automatically once this transaction is confirmed.

Such a configuration would firstly increase the speed of the entire settlement cycle from days to minutes or even seconds, where we would essentially have continuous settlement. There are a number of industry initiatives already in the digital asset transfer and settlement space, and we mention indicatively R3 CEV\footnote{\url{r3cev.com}}, Digital asset holdings\footnote{\url{digitalasset.com}}, Symbiont\footnote{\url{symbiont.io}}, Chain\footnote{\url{chain.com}} and SETL\footnote{\url{setl.io}}. HitFin\footnote{\url{www.hitfin.com}} has proposed an alternative approach from the one described here, where trades are cleared bilaterally on a private blockchain, in less than 17 seconds. Besides the fast transaction settlement and automatic settlement of contracts upon maturity, all reporting, compliance and collateral management can be handled through the blockchain, thus reducing backoffice coss.  

\begin{figure}
\includegraphics[width = \textwidth]{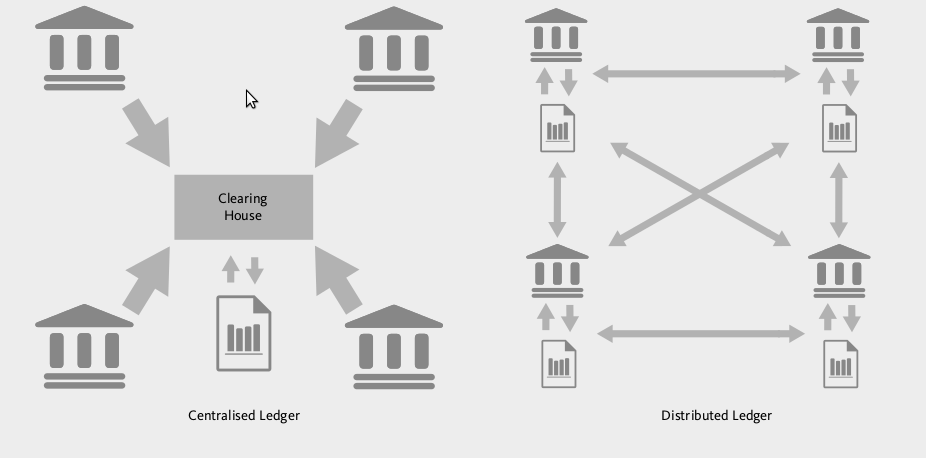}
\caption{Clearing in a centralised and decentralised ledger. Image source: The Fintech 2.0 Paper: rebooting financial services, available at \url{http://www.finextra.com/finextra-downloads/newsdocs/The\%20Fintech\%202\%200\%20Paper.PDF}. In this instance, the blockchain version of clearing is a complete decentralisation where the roles of the exchange, the clearing house and the settlement system are no longer separated since they are combined into a single distributed blockchain architecture. However, once could argue that for reasons governance it may be more appropriate to consider intermediate architectures for blockchain decentralisation that still incorporates these separations of roles, such as the ones discussed in this section.}
\end{figure}
 
%%%%%%%%%%%%%%%%%%%%%%%%%%%%%%%%%%%%%%%%%%%%%%%%%%%%%%%%
\section{Blockchain Technology and Multi-signature Escrow Services}
%%%%%%%%%%%%%%%%%%%%%%%%%%%%%%%%%%%%%%%%%%%%%%%%%%%%%%%%
%
To complete discussions on blockchain technologies and their applications in banking settings we mention some brief details on their influence in multi-signature Escrow services. This could be particularly beneficial for the banking ledger and settlement process applications described above as a means of resolving disputed financial transactions. As has been discussed above, some versions of blockchain technologies such as the version created for the bitcoin protocol require that the transactions are irreversible. This can be both beneficial and also a hinderance to the application of such technologies. Due to this feature, there is no aspect of dispute mediation present, as there would be with other electronic based payment systems. It has been reported, see Chapter 4 of \cite{franco2014understanding}, that as a result of this feature, the transaction processing costs of bitcoin based payment services are significantly lower than those of other remittance market services. Whilst the average fee for typical remittance market services is around 8\%-9\%, those currently available for Bitcoin payment services are only of the order of 0.01\%-0.05\%, largely due to the lower cost of not needing to process or perform disputes in transactions. However, for more general applications, this lack of reversibility can be problematic, see discussions in \cite{buterin2014multisig}. Certainly, in many blockchain technology applications it may be beneficial to have an ability to make modifications or amendments to disputed transactions. 

Ideas to achieve this have begun to be explored in settings of blockchain technology such as for payment or contract transfers involving multiple signatories via an Escrow service. An Escrow service will provide an opportunity to perform a dispute resolution between two transactions in the following manner. The two transacting parties identify a trusted intermediary party (Escrow service) who then provides them each with an electronic ``address'' and maintains a private cryptographic key to control this address. Then to perform the transactions, agent A sends the information/e-property/e-money to a 2-of-3 multi-signature address. The addresses involved are those of the other party (agent B), the agent (A) and the Escrow service provider, then one has three possible outcomes to consider:
\begin{itemize}
\item The agent A may receive the information/e-property/e-money as agreed and in return they verify a transaction that releases the renumeration or alert of receipt to agent B through the 2-of-3 multisignature address to the Escrow agents address. The Escrow agent then signs the 2-of-3 multisignature address with their own key and publishes it on the blockchain.;
\item If there is a problem with the contractual obligations agreed from agent A's perspective and agent B agrees with this, then agent B would sign the 2-of-3 multisignature address and send appropriate renumeration to agent A's address. Then Agent A upon receipt of this signs the 2-of-3 multisignature address and publishes it on the blockchain.
\item If there is a dispute between agent A and agent B as to the contractual agreements and who should be renumerated, then the Escrow service provides an intervention. Here the Escrow service decides the outcome of the funds by signing the 2-of-3 multisignature address renumerating the appropriate party and then the party receiving renumeration signs the 2-of-3 multisignature address and publishes this on the blockchain. Note, in this third case the Escrow service would typically charge a fee.
\end{itemize}
We believe such approaches can be more widely abopted in blockchain technologies via automated smart contract structures.

%%%%%%%%%%%%%%%%%%%%%%%%%%%%%%%%%%%%%%%%%%%%%%%%%%%%%%%%
%%%%%%%%%%%%%%%%%%%%%%%%%%%%%%%%%%%%%%%%%%%%%%%%%%%%%%%%
\section{Conclusions}
\label{sec:conc}
%%%%%%%%%%%%%%%%%%%%%%%%%%%%%%%%%%%%%%%%%%%%%%%%%%%%%%%%
%%%%%%%%%%%%%%%%%%%%%%%%%%%%%%%%%%%%%%%%%%%%%%%%%%%%%%%%

This chapter has served to first highlight some of the recent innovations in the space of blockchain technologies. It has outlined some important aspects of blockchain architectures and their commonality and distinguishing features from different types of database structures. It has then described a number of features that are vital from a financial application perspective, including permissioning, data integrity, data security and data authenticity as well as important regulatory requirements relating to account provisioning for financial asset reporting, and the blockchain aspects that can help adhere to these. Then several innovative new areas of development for second generation blockchain techonologies are detailed, including central bank treasury ledgers, retail and investment bank ledgers, trading, settlement and clearing processes, finishing with a discussion on multi-signature Escrow services. Like all prior disruptive technologies there will be beneficial and detrimental aspects of blockchain technologies that will need to be carefully considered prior to development and commercialisation of the ideas presented in this chapter. However, we believe that with the onset of the internet of money, the blockchain revolution will play an integral part in this brave new world.

\bibliographystyle{authordate1}
\bibliography{bit,allbit}

\end{document}